\documentstyle[preprint,pra,aps,epsf]{revtex}
\begin{document}
\tighten

\title{Using a quantum computer to investigate quantum chaos}

\author{R\"udiger Schack\thanks{Email: r.schack@rhbnc.ac.uk}}
\address{Department of Mathematics, Royal Holloway,
University of London \\ Egham, Surrey TW20 0EX, UK}
\date{\today}
\maketitle

\begin{abstract}
We show that the quantum baker's map, a prototypical map invented for
theoretical studies of quantum chaos, has a very simple realization in terms of
quantum gates. Chaos in the quantum baker's map could be investigated
experimentally on a quantum computer based on only 3 qubits.
\end{abstract}

\vskip20mm

Since the discovery that a quantum computer can in principle factor large
integers in polynomial time\cite{Shor1994,Ekert1996}, quantum information has
become a major theoretical and experimental research topic, focusing on
properties, applications, generation and preservation of highly entangled
quantum states \cite{recent}. Although it is not
clear if a full-scale quantum computer will ever be
realized\cite{Unruh1995,Plenio1996}, experiments with quantum gates are being
performed at present\cite{Cirac1995,Monroe1995a,Gershenfeld1997,Cory1997}. It is
important to devise applications for early quantum computers which are
incapable of doing large-scale computations like factoring.

Early quantum computers appear to
be well suited to study the quantum dynamics of simple quantum maps.  The
quantum baker's map\cite{Balazs1989}, one of the simplest quantum maps used in
quantum chaos research, has been extensively studied in recent years
\cite{Saraceno1990,Saraceno1992a94,Schack1993a96,Lakshminarayan1993a,%
Lakshminarayan1995,DaLuz1995}.  Up to now, it has been regarded
as a purely theoretical toy model. As a consequence of recent progress in the
field of quantum
computing\cite{Cirac1995,Monroe1995a,Gershenfeld1997,Cory1997}, however, an
experimental realization of the quantum baker's map seems possible in the very
near future.

Any unitary operator can be approximated by a sequence of simple quantum
gates\cite{Deutsch1985,Barenco1995e,DiVincenzo1995}. The main result of this
paper is that the quantum baker's map has a particularly simple realization in
terms of quantum gates. The quantum baker's map displays behavior of
fundamental interest even for a small dimension of Hilbert space. Numerical
simulations\cite{Schack1993a96} in $D=16$ dimensional Hilbert space suggest
that a rudimentary quantum computer based on as few as three qubits (i.e. three
two-state systems spanning $D=8$ dimensional Hilbert space) could be used to
study chaos in the quantum baker's map. In particular, it may be possible to
find experimental evidence for hypersensitivity to perturbation, a proposed
information-theretical characterization of quantum
chaos\cite{Schack1993a96,Caves1993b,Schack1992a96,Schack1994b}.

The classical baker's transformation\cite{Arnold1968} maps the unit square $0
\leq q,p \leq 1$ onto itself according to
\begin{equation}
(q,p) \longmapsto \left\{  \begin{array}{ll}
\Bigl(2q,{1\over2}p\Bigr)\;,       &\mbox{if $0\leq q\leq{1\over2}$,} \\
\Bigl(2q-1,{1\over2}(p+1)\Bigr)\;, &\mbox{if ${1\over2}<q\leq1$.}
                           \end{array}  \right. 
\label{eqcbaker}
\end{equation}
This corresponds to compressing the unit square in the $p$ direction and
stretching it in the $q$ direction, while preserving the area, then cutting it
vertically, and finally stacking the right part on top of the left part---in
analogy to the way a baker kneads dough.  

To define the quantum baker's map\cite{Balazs1989}, we quantize the unit square
following \cite{Saraceno1990,Weyl1950}. To represent the unit square
in $D$-dimensional Hilbert space, we start with unitary ``displacement''
operators $\hat U$ and $\hat V$, which produce displacements in the
``momentum'' and ``position'' directions, respectively, and which obey the
commutation relation \cite{Weyl1950}
\begin{equation}
\hat U\hat V = \hat V\hat U\epsilon \;,
\end{equation}
where $\epsilon^D=1$. We choose $\epsilon=e^{2\pi i/D}$.  We further assume
that $\hat V^D=\hat U^D=1$, i.e., periodic boundary conditions. It
follows\cite{Saraceno1990,Weyl1950} that the operators $\hat U$ and $\hat V$
can be written as
\begin{equation}
\hat U=e^{2\pi i\hat q}\qquad\mbox{and}\qquad \hat V=e^{-2\pi i\hat p} \;.
\end{equation}
The ``position'' and ``momentum'' operators $\hat q$ and $\hat p$ both have 
eigenvalues $j/D$, $j=0,\ldots,D-1$.

In the following, we restrict the discussion to the case $D=2^L$, i.e., the
dimension of Hilbert space is a power of two.  For consistency
of units, let the quantum scale on ``phase space'' be $2\pi\hbar=1/D=2^{-L}$. A
transformation between the position basis $\{|q_j\rangle\}$ and the momentum
basis $\{|p_j\rangle\}$ is effected by the discrete Fourier transform $F_L'$,
defined by the matrix elements
\begin{equation}
(F_L')_{kj} = \langle p_k|q_j\rangle = 
\sqrt{2\pi\hbar}\; e^{-ip_kq_j/\hbar} =
{1\over\sqrt D}e^{-2\pi ikj/D}\;.
\label{eqfourier}
\end{equation}

There is no unique way to quantize a classical map. Here we adopt the quantized
baker's map introduced by Balazs and Voros \cite{Balazs1989} and defined by the
matrix
\begin{equation}
T' = F_L'^{-1} \left( \begin{array}{cc}
                       F_{L-1}' & 0 \\ 0 & F_{L-1}'
                       \end{array} \right) \;,
\label{eqqbaker}
\end{equation}
where the matrix elements are to be understood relative to the position basis
$\{|q_j\rangle\}$. Saraceno {\cite{Saraceno1990}} has introduced a quantum
baker's map with stronger symmetry properties by using antiperiodic boundary
conditions, but in this article we restrict the discussion to periodic boundary
conditions as used in\cite{Balazs1989}.

The discrete Fourier transform used in the definition of the quantum baker's
map (\ref{eqqbaker}) plays a crucial role in quantum computation and can be
easily realized as a quantum network using simple quantum gates.  The following
discussion of the quantum Fourier transform closely
follows\cite{Ekert1996}. The $D=2^L$ dimensional Hilbert space modeling the
unit square can be realized as the product space of $L$ qubits (i.e. $L$
two-state systems) in such a way that
\begin{equation}
|q_j\rangle =
|j_{L-1}\rangle\otimes|j_{L-2}\rangle\otimes\cdots\otimes|j_0\rangle \;,
\label{eqtensor}
\end{equation}
where $j=\sum j_k2^k$, $j_k\in\{0,1\}$ ($k=0,\ldots L-1$), 
and where each qubit has basis states $|0\rangle$ and $|1\rangle$.

To construct the quantum Fourier transform, two basic unitary operations or 
{\it quantum gates\/} are needed: the gate $A_m$ acting on the $m$th qubit and
defined in the basis $\{|0\rangle,|1\rangle\}$ by the matrix
\begin{equation}
A_m= {1\over\sqrt2}
      \left( \begin{array}{cc}
       1 & 1 \\ 1 & -1        \end{array} \right)  \;,
\end{equation}
and the gate $B_{mn}$ operating on the $m$th and $n$th qubits and defined by
\begin{equation}
B_{mn}\, |j_{L-1}\rangle\otimes\cdots\otimes|j_0\rangle =
  e^{i\phi_{mn}}\, |j_{L-1}\rangle\otimes\cdots\otimes|j_0\rangle \;,
\end{equation}
where 
\begin{equation}
\phi_{mn} = \left\{     \begin{array}{ll}
\pi/2^{n-m} & \mbox{if $j_m=j_n=1$ ,} \\
0           & \mbox{otherwise.}
                     \end{array}
         \right. 
\end{equation}
In addition we define the gate $S_{mn}$ which swaps the qubits $m$ and $n$.

The discrete Fourier transform $F_L$ can now be expressed in terms of the three
types of gates as
\begin{eqnarray}
F_L & = & S\times \big(A_0 B_{01} \cdots B_{0,L-1}\big) \times\cdots \nonumber\\
    &&  \times \big(A_{L-3}B_{L-3,L-2}B_{L-3,L-1}\big) \\
    &&  \times \big(A_{L-2}B_{L-2,L-1}\big)  \times \big(A_{L-1}\big) \nonumber
\end{eqnarray}
where
\begin{equation}
S = \left\{     \begin{array}{ll}
  S_{0,L-1}S_{1,L-2}\cdots S_{L/2-1,L/2} & \mbox{for $L$ even,} \\
  S_{0,L-1}S_{1,L-2}\cdots S_{(L-3)/2,(L+1)/2} & \mbox{for $L$ odd,}
                     \end{array}
         \right. 
\end{equation}
reverses the order of the qubits.
The quantum baker's map (\ref{eqqbaker}) is then given by
\begin{equation}
T = F_L^{-1} \big(I\otimes F_{L-1}\big) \;,
\label{eqgbaker}
\end{equation}
where $F_{L-1}$ acts on the $L-1$ least significant qubits, and $I$ is the
identity operator acting on the most significant qubit.  The gates
corresponding to the bit-reversal operator $S$ can be saved if the qubits in
the tensor product (\ref{eqtensor}) are relabeled after each execution of $F_L$
or $F_{L-1}$.

In $D=8=2^3$ dimensional Hilbert space, one iteration of the quantum baker's
map is performed by the short sequence of gates
\begin{equation}
T = S_{02} A_0 B_{01}^\dagger B_{02}^\dagger A_1 B_{12}^\dagger A_2 S_{01} A_0
B_{01} A_1 \;.
\end{equation}

This implementation of the quantum baker's map can be viewed in two
complementary ways. On the one hand, it shows that the quantum baker's map can
be efficiently simulated on a quantum computer. A 30-qubit quantum
computer could perform simulations that are
virtually impossible on present-day classical computers.

On the other hand, an iteration of the gate sequence (\ref{eqgbaker}) on an
$L$-qubit quantum computer is a physical realization of the quantum baker's
map. This opens up the possibility of an experimental investigation of chaos in
a physical system in a purely quantum regime.


\begin{thebibliography}{10}

\bibitem{Shor1994}
P.~W. Shor,  in {\em Proceedings of the 35th Annual Symposium on the Theory of
  Computer Science}, edited by S. Goldwasser (IEEE Computer Society Press, Los
  Alamitos, California, 1994), p.\ 124.

\bibitem{Ekert1996}
A. Ekert and R. Jozsa, Rev.\ Mod.\ Phys.\ {\bf 68},  733  (1996).

\bibitem{recent} See recent issues of PRL and PRA and the e-print archive
quant-ph.

\bibitem{Unruh1995}
W.~G. Unruh, Phys.\ Rev.\ A {\bf 51},  992  (1995).

\bibitem{Plenio1996}
M.~B. Plenio and P.~L. Knight, Phys.\ Rev.\ A {\bf 53},  2986  (1996).

\bibitem{Cirac1995}
J.~I. Cirac and P. Zoller, Phys.\ Rev.\ Lett.\ {\bf 74},  4091  (1995).

\bibitem{Monroe1995a}
C. Monroe {\it et~al.}, Phys.\ Rev.\ Lett.\ {\bf 75},  4714  (1995).

\bibitem{Gershenfeld1997}
N.~A. Gershenfeld and I.~L. Chuang, Science {\bf 275},  350  (1997).

\bibitem{Cory1997}
D. Cory, A. Fahmy, and T. Havel, Proc.\ Nat.\ Acad.\ Sci.\ USA {\bf 94},  1634
  (1997).

\bibitem{Balazs1989}
N.~L. Balazs and A. Voros, Ann.\ Phys.\ {\bf 190},  1  (1989).

\bibitem{Saraceno1990}
M. Saraceno, Ann.\ Phys.\ {\bf 199},  37  (1990).

\bibitem{Saraceno1992a94} 
M. Saraceno and A. Voros, Chaos {\bf 2}, 99 (1992);
Physica D {\bf 79}, 206 (1994).

\bibitem{Schack1993a96}
R. Schack and C.~M. Caves, Phys.\ Rev.\ Lett.\ {\bf 71},  525  (1993);
Phys.\ Rev.\ E {\bf 53},  3257  (1996).

\bibitem{Lakshminarayan1993a}
A. Lakshminarayan and N.~L. Balazs, Ann.\ Phys.\ {\bf 226},  350  (1993).

\bibitem{Lakshminarayan1995}
A. Lakshminarayan, Ann.\ Phys.\ {\bf 239},  272  (1995).

\bibitem{DaLuz1995}
M.~G.~E. da~Luz and A.~M. Ozorio~de Almeida, Nonlinearity {\bf 8},  43  (1995).

\bibitem{Deutsch1985}
D. Deutsch, Proc.\ R. Soc.\ Lond.\ A {\bf 400},  97  (1985).

\bibitem{Barenco1995e}
A. Barenco {\it et~al.}, Phys.\ Rev.\ A {\bf 52},  3457  (1995).

\bibitem{DiVincenzo1995}
D.~P. Di~Vincenzo, Phys.\ Rev.\ A {\bf 51},  1015  (1995).

\bibitem{Caves1993b}
C.~M. Caves,  in {\em Physical Origins of Time Asymmetry}, edited by J.~J.
  Halliwell, J. P\'erez-Mercader, and W.~H. Zurek (Cambridge University Press,
  Cambridge, England, 1993), p.\ 47.

\bibitem{Schack1992a96}
R. Schack and C.~M. Caves, Phys.\ Rev.\ Lett.\ {\bf 69},  3413  (1992);
Phys.\ Rev.\ E {\bf 53},  3387  (1996).

\bibitem{Schack1994b}
R. Schack, G.~M. D'Ariano, and C.~M. Caves, Phys.\ Rev.\ E {\bf 50},  972
  (1994).

\bibitem{Arnold1968}
V.~I. Arnold and A. Avez, {\em Ergodic Problems of Classical Mechanics}
  (Benjamin, New York, 1968).

\bibitem{Weyl1950}
H. Weyl, {\em The Theory of Groups and Quantum Mechanics} (Dover, New York,
  1950).

\end{thebibliography}

\end{document}